\documentclass{aa}  

\usepackage{bm}

\usepackage{natbib}
\bibpunct{(}{)}{;}{a}{}{,} 

\usepackage{graphicx}
\usepackage{txfonts}
\usepackage{lipsum}
\usepackage{subcaption}         
\usepackage{lscape}             
\usepackage{placeins}           
                                
\begin{document}

   \title{Solar Wind Proton Heating and its Effect on Temperature Anisotropy Evolution between 0.05 and 1 au}


%
%
%

\author{
Etienne Berriot\inst{1,2}\corrauth{etienne.berriot@obspm.fr}
\and Arnaud Zaslavsky\inst{1}\email{arnaud.zaslavsky@obspm.fr}
\and Olga Alexandrova\inst{1}\email{olga.alexandrova@obspm.fr}
\and Pascal D\'emoulin\inst{1}\email{pascal.demoulin@obspm.fr}
\and Petr Hellinger\inst{3,4}\email{petr.hellinger@asu.cas.cz}
\and Lorenzo Matteini\inst{5}\email{l.matteini@imperial.ac.uk}
}

\institute{
LIRA, Observatoire de Paris, Universit\'e PSL, Sorbonne Universit\'e, Universit\'e Paris Cit\'e, CY Cergy Paris Universit\'e, CNRS, 92190 Meudon, France
\and Dipartimento di Fisica, Universit\`a della Calabria, Rende, I-87036, Italy
\and Astronomical Institute of the Czech Academy of Sciences, Prague, Czechia
\and Institute of Atmospheric Physics of the Czech Academy of Sciences, Prague, Czechia
\and The Blackett Laboratory, Department of Physics, Imperial College, London SW7 2AZ, UK
}

\date{\today}

 
\abstract{
This study focuses on the radial evolution of the solar wind proton adiabatic invariants and temperature anisotropies in the inner heliosphere.
More specifically, we study in-situ measurements provided by the Parker Solar Probe, between 0.05~au and 0.25~au from the Sun, and Solar Orbiter spacecraft between 0.3~au and 1~au.
Throughout the studied range of radial distances, we observe a significant average heating in the direction perpendicular to the local magnetic field for both fast and slow solar wind populations.
On the other hand, there is no clear deviation from adiabaticity in the parallel direction regardless of the wind speed.
The perpendicular heating is enough to significantly reduce the generation of the temperature anisotropy expected from a double adiabatic evolution.
Despite the heating, an important portion of the solar wind (especially the slower wind streams) develops substantial anisotropies with higher parallel temperatures, which eventually become constrained by kinetic firehose instabilities.
}


\maketitle


\nolinenumbers 

\section{Introduction} \label{sec:intro}

The solar wind is a magnetized and weakly collisional medium, which allows particle velocity distribution functions to develop complex features such as anisotropies, agyrotropies, beams and velocity drifts between the different species \citep[see e.g.,][]{Feldman_1975_SW_electrons, Marsch_1982_Helios_protons_evolution, Stverak_2009_electrons_suprathermal_radial_evol, Marsch_2012_Helios_evol_distrib_func}.
The solar wind expansion then largely deviates from that of a collisional gas with isotropic Maxwellian particles velocity distributions, and naturally leads to the formation of temperature anisotropies.
These anisotropies and non-Maxwellian features are however constrained by a variety of kinetic processes, such as instabilities, which bring back the plasma closer to thermal equilibrium \citep{Gary_1993_instabilities_book, Matteini_2012_ion_kinetics_evol_SW, Verscharen_2022_SW_electron_instabilities}.
In-situ observations at 1~au moreover show that the global distribution of solar wind protons measurements in the $\beta_{p \parallel} \, \textit{-} \, T_{p \perp} / T_{p \parallel}$ plane is reasonably well constrained by fitted marginal stability conditions of linear kinetic instabilities \citep{Hellinger_2006_anisotropy_thresholds_SW}.
Here, and in the following of this letter, $\beta_{p \parallel} = n_p k_B T_{p \parallel} / (B^2 / 2 \mu_0)$ denotes the parallel proton plasma beta, with $n_p$ the proton density, while $T_{p \perp}$ and $T_{p \parallel}$ are the perpendicular and parallel proton temperatures with respect to the local magnetic field $\bm{B}$ of magnitude $B$.

Previous studies have identified a perpendicular proton heating (as compared to an adiabatic expansion) in Helios measurements between 0.3 and 1~au, while neither heating nor cooling was observed in the parallel direction \citep{Marsch_1983_adiabatic_invariants_Helios, Perrone_2019_FSW_Helios_adiab_inv, Zaslavsky_2023_SW_heating_rate_evaluation}. 
On the other hand, hybrid-PIC expanding box simulations by \citet{Matteini_2006_parallel_FH_HEB, Hellinger_2019_3D_FH_HEB} have shown the plasma distribution in the $\beta_{p \parallel} \, \textit{-} \, T_{p \perp} / T_{p \parallel}$ plane getting closer to the kinetic firehose marginal stability conditions, and then getting regulated when exceeding it.
A qualitatively similar evolution in fast solar wind observations between 0.3 and 2.5~au was reported by \citet{Matteini_2007_anisotropy_Helios-Ulysses}.
The authors observed a development of the proton temperature anisotropy with distance, slower than the adiabatic prediction (in agreement with a perpendicular heating), which was then seemingly constrained by the firehose instability.
In addition, \citet{Jagarlamudi_2025_SolO_T_anisotropies} have studied Solar Orbiter measurements (between 0.3 and 1~au) of proton and alpha particles, each fitted by bi-Maxwellian velocity distributions functions.
Their results indicate a larger temperature anisotropy ratio $T_{p \perp} / T_{p \parallel}$ in the fast wind (with speed superior to 600 km/s) than in the slow one (with speed inferior to 400 km/s), as well as a decreasing $T_{p \perp} / T_{p \parallel}$ with solar distance for both fast and slow wind.
Similar results were also obtained by \citet{Marsch_1982_Helios_protons_evolution} using Helios data between 0.3 and 1 au.

The recently launched Parker Solar Probe \cite[PSP,][]{Fox_2016_PSP} mission provides in-situ solar wind measurements closer to the Sun than ever, thus enabling the study of heating and temperature anisotropy evolution in the young solar wind.
In particular, \citet{Mozer_2023_PSP_heating_rates} provides the first study of non-adiabatic solar wind proton heating accounting for temperature anisotropies using PSP observations.
Using measurements from encounters 6-9 and going from 20 to 160 solar radii ($R_s$), so from 0.1 to 0.75~au, the authors reported an average heating in the perpendicular direction and estimated the associated heating rate, but did not observe any deviation from adiabaticity in the parallel direction.
A statistical study of several solar wind parameters radial evolution using PSP measurements during encounters 1-24 was moreover carried by \citet{Yogesh_2026_PSP_radial_evolution}.
Notably, between 0.05 and 0.15~au (10 and 30~$R_s$), the authors found an overall decrease of $T_{p \perp} / T_{p \parallel}$ with distance from the Sun, with an evolution weaker than the adiabatic prediction.
Accordingly, using PSP data from encounter 1-21 (also between 0.05 and 0.15~au and not sorted by wind speed), \citet{Coello-Guzman_2026_anisotropy_Brazil_PSP} have found an average radial evolution of $T_{p \perp} / T_{p \parallel}$ and of the plasma distribution in the $\beta_{p \parallel} \, \textit{-} \, T_{p \perp} / T_{p \parallel}$ plane different from the double CGL prediction, and consistent with a perpendicular heating.
Nonetheless, studies of PSP's measurements within 0.15~au by \citet{Short_2024_PSP_Brazil, Coello-Guzman_2026_anisotropy_Brazil_PSP} still indicate that the overall distribution of the solar wind plasma evolves towards regions of lower $T_{p \perp} / T_{p \parallel}$ and higher $\beta_{p \parallel}$ with distance from the Sun, and eventually becomes constrained by the kinetic firehose marginal stability conditions.
This behavior was also observed by \citet{Payne_2026_non-adiaba_SBs_PSP} using PSP measurements of both solar wind with speed $v < 300$~km/s and $v > 300$~km/s separately, and at distances between 0.05 and 0.25~au (10 and 50~$R_s$).

In this paper we investigate, for both the slow and fast solar wind separately, radial profiles of the proton CGL \citep{Chew_1956_CGL_invariants} adiabatic invariants and the evolution of the proton temperature anisotropy, at distances $R \in [0.05, 0.25]$~au from the Sun for PSP encounters 1-25 (from November 2018 to September 2025) and $R \in [0.3, 1]$~au from the Sun for SolO measurements from June 2020 to November 2025.
Section \ref{sec:theory-equations} summarizes the theoretical background with a few key equations.
In Section \ref{sec:non-adiab} we show, through monitoring of the adiabatic invariants, that there is no evident average parallel proton heating or cooling regardless of the solar wind speed, while there is a significant perpendicular proton heating with distance for both the slower and faster solar winds.
We then estimate the average non-adiabatic perpendicular proton heating rates, which decrease with distance from the Sun, and are always higher in the faster solar wind.
Section \ref{sec:anisotropy_evolution} then focuses on the evolution of proton temperature anisotropy and plasma distribution in the $\beta_{p \parallel} \, \textit{-} \, T_{p \perp} / T_{p \parallel}$ plane.
For the slower and faster winds, we observe distinct behaviors that are different from the adiabatic predictions and consistent with a perpendicular heating.
Finally, we discuss our results and conclude in Section \ref{sec:conclusion}.

\section{Energy Evolution Equations and Adiabatic CGL Predictions} \label{sec:theory-equations}

This study focuses on the solar wind protons (subscript $p$) internal energy evolution.
Integration of the proton Boltzmann equation second velocity moment leads to the pressure tensor evolution equation.
When projected along the directions parallel and orthogonal to the magnetic field (subscript $\parallel$ and $\perp$, respectively), this provides the evolution of $C_{p \parallel}$ and $C_{p \perp}$ (see Appendix \ref{sec:appendix_energy_equations}), defined as
\begin{align}
      C_{p \parallel} = T_{p \parallel} \left( \frac{B}{n_p} \right)^2 \, ,
      &&  C_{p \perp} = T_{p \perp} / B \, ,
\end{align}
with $T_{p \parallel}$ and $T_{p \perp}$ the parallel and perpendicular proton temperatures, respectively, $n_p$ the proton density, and $B$ the magnetic field magnitude.
The aforementioned evolution equation can be written in a simple form (see \citet{Hunana_2019_CGL_intro,Zaslavsky_2023_SW_heating_rate_evaluation} and Appendix \ref{sec:appendix_energy_equations} for more details)
\begin{align}
    \frac{1}{2} \frac{\mathrm{d}}{\mathrm{d}t} \ln C_{p \parallel} &= Q_{p \parallel} / P_{p \parallel} \, , \label{eq:C_para_evol} \\
    \frac{\mathrm{d}}{\mathrm{d}t} \ln C_{p \perp} &= Q_{p \perp} / P_{p \perp} \, , \label{eq:C_perp_evol}
\end{align}
where $\mathrm{d} / \mathrm{d} t = \partial_t + \bm{u}_p \cdot \bm{\nabla}$ is the total temporal proton derivative with $\bm{u}_p$ the proton bulk velocity, while $P_{p \parallel}$ and $P_{p \perp}$ are the parallel and perpendicular proton pressures, respectively.
The quantities $Q_{p \perp}$ and $Q_{p \parallel}$ are the proton non-adiabatic heating rates densities.

In the absence of collisions, heat fluxes, non-ideal electric fields ($\mathbf{E} \neq -\mathbf{u} \times \mathbf{B}$), and contributions from the non-gyrotropic part of the pressure tensor, the plasma is behaving adiabatically and both $C_{p \parallel}$ and $C_{p \perp}$ are conserved.
Then $Q_{p \perp} = Q_{p \parallel} = 0$, so Equations (\ref{eq:C_para_evol}) and (\ref{eq:C_perp_evol}) reduce to the usual double-CGL \citep{Chew_1956_CGL_invariants} approximation
\begin{align}
    \frac{\mathrm{d}}{\mathrm{d}t} C_{p \parallel} &= 0 \, , \label{eq:C_para_adiab} \\
    \frac{\mathrm{d}}{\mathrm{d}t} C_{p \perp} &= 0 \, . \label{eq:C_perp_adiab}
\end{align}
Therefore, $C_{p \parallel}$ and $C_{p \perp}$ are usually referred to as the parallel and perpendicular adiabatic invariants, respectively.

Conservation of the adiabatic invariants then implies
\begin{align}
      T_{p \parallel} \propto \left( \frac{n_p}{B} \right)^2 \, ,
      && T_{p \perp} \propto B \, ,
\end{align}
so the proton temperature anisotropy and parallel plasma beta evolve as
\begin{align}
      \frac{T_{p \perp}}{T_{p \parallel}} \propto \frac{B^3}{n_p^2} \, , && \beta_{p \parallel} \propto \frac{n_p^3}{B^4} \, . \label{eq:anisotropy-beta_evol_adiab}
\end{align}

Assuming then a steady-state solar wind spherically expanding at constant speed ($n_p \propto R^{-2}$), with a radial magnetic field ($|\mathbf{B}| = | B_R | \propto R^{-2}$), then gives:
\begin{align}
      \frac{T_{p \perp}}{T_{p \parallel}} \propto R^{-2} \, , && \beta_{p \parallel} \propto R^{2} \, . \label{eq:anisotropy-beta_evol_adiab_R}
\end{align}
Thus, for an adiabatic solar wind expansion close to the Sun (where $\mathbf{B}$ is mostly radial), the temperature anisotropy and parallel beta both increase with $R$.
This then brings the plasma distribution towards the kinetic firehose instability conditions.
Once the instability is triggered, important fluctuations are produced around the ion kinetic scales, decreasing the temperature anisotropy and inducing a non-steady and non-adiabatic behavior of the plasma, which thus render the adiabatic scalings no longer valid \citep{Hunana_2017_CGL_myth,Hellinger_2019_3D_FH_HEB}.
Hence, even in the absence of external heating, the solar wind cannot maintain a steady adiabatic expansion close to the Sun.
This evolution can be further modified under the effects of other non-adiabatic processes, such as turbulent heating for example.
Then, studying the evolution of $C_{p \parallel}$, $C_{p \perp}$ and temperature anisotropy provides observational constraints on which physical processes dominate during the solar wind evolution.

In this study, we consider PSP measurements taken at distances $R \in [0.05, 0.25]$~au for encounters 1 to 25 (from November 2018 to September 2025), and SolO measurements at $R \in [0.05, 0.25]$~au between July 2020 and November 2025.
We analyze integrated moments of the proton velocity distribution functions from measurements provided by the SPAN-Ion electrostatic analyzer \citep{Livi_2022_SPI} part of the SWEAP \citep{Kasper_2016_SWEAP} suite for PSP, and those provided by the instrument PAS from the SWA suite \citep{Owen_2020_SWA} for SolO.
For PSP, the proton densities have been considered to be the same as the electron density estimated through quasi-thermal noise (QTN) spectroscopy \citep{Meyer-Vernet_2017_QTN_review, Moncuquet_2020_QTN_PSP}, in order to get more robust systematic estimations because SPAN-Ion measured densities can suffer from the instrument limited field-of-view (FOV).
We also use magnetic field measurements given by the fluxgate magnetometer from the FIELDS instrument suite \citep{Bale_2016_FIELDS} on PSP, and by the MAG magnetometer \citep{Horbury_2020_MAG} on SolO.
We excluded PSP measurements from 2023-03-13 05:00:00 UTC to 2023-03-14 03:00:00 UTC because of temperature anisotropy measurements being strongly affected around an interplanetary coronal mass ejection (ICME) crossing.
Other ICMEs have been removed, from both PSP and SolO data, using the HELIO4CAST ICME catalog\footnote{The catalog used here is publicly available at \url{https://doi.org/10.6084/m9.figshare.6356420.v23}, see also \url{https://helioforecast.space/icmecat}.} \citep{Moestl_2026_ICMEs_evolution_catalog}.
We note that the removal of ICMEs mostly lowers the measurements dispersion and do not qualitatively affect the results presented below.
PSP's encounter 11 has moreover not been considered due to the lack of QTN measurements.

\section{Non-Adiabatic Solar Wind Expansion} \label{sec:non-adiab}

We are first interested in checking if measurements by PSP and SolO indicate, or not, an adiabatic expansion of the solar wind protons.
Assuming a stationary and radial expanding solar wind, the total temporal proton derivatives reduce to $\mathrm{d}/\mathrm{d}t = u_{p, R} \, \partial_R$, where $u_{p, R}$ is the proton's radial speed.
Then, according to Equations (\ref{eq:C_para_adiab}) and (\ref{eq:C_perp_adiab}), deviations from adiabaticity can directly be identified by monitoring the $C_{p (\perp, \parallel)}$ invariants evolution with distance \citep{Marsch_1983_adiabatic_invariants_Helios, Zaslavsky_2023_SW_heating_rate_evaluation}.
This also allows to quantify the heating and/or cooling rates associated to a non-adiabatic proton behavior using Equations (\ref{eq:C_para_evol}) and (\ref{eq:C_perp_evol}).

\begin{figure}[ht!]
\centering
      \includegraphics[width=\hsize]{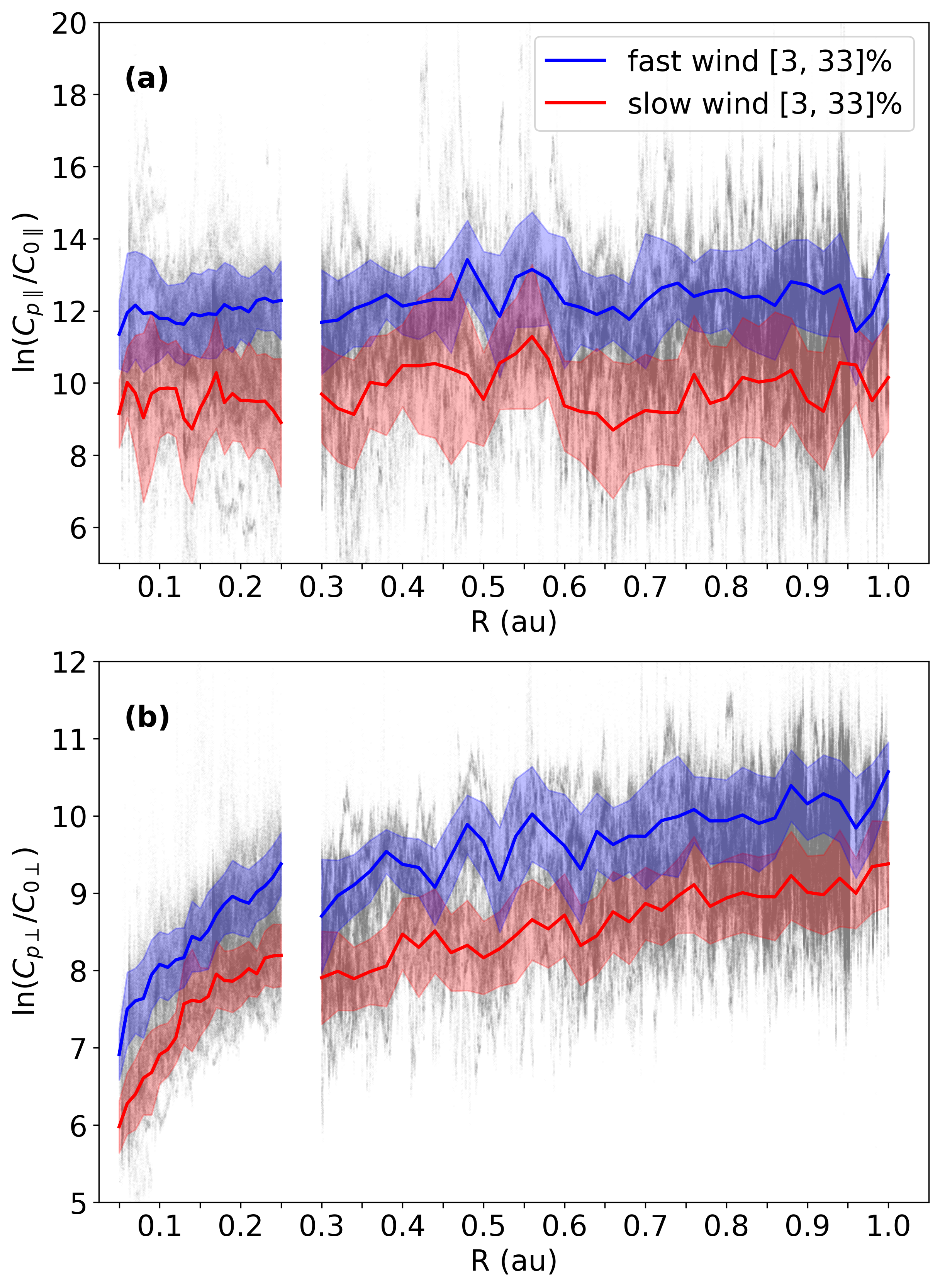}
      \caption{
            Radial evolution of the proton adiabatic invariants' logarithm $\ln C_{p \parallel}$ (a) and $\ln C_{p \perp}$ (b) measured by PSP for $R \in [0.05, 0.25]$~au and SolO for $R \in [0.3, 1]$~au.
            The invariants are normalized by values $C_{0 \parallel} = 1 $~K.nT$^{2}$.cm$^{6}$ and $C_{0 \perp} = 1 $~K.nT$^{-1}$.
            Grey dots are measurements averaged over a 1~min time interval.
            Blue and red colors are associated to faster and slower solar wind speeds.
            Plain curves are averages over radial bins of $10^{-2}$~au for PSP data $2 \times 10^{-2}$~au for SolO data; the standard deviation of the two distributions are overlaid in semi-transparent colors.
      }
      \label{fig:adiabatic_invariants} 
\end{figure}

Figure \ref{fig:adiabatic_invariants} shows the radial evolution of both proton adiabatic invariants logarithm: $\ln C_{p \parallel}$ in panel (a) and $ \ln C_{p \perp}$ in panel (b).
Measurements, displayed as grey dots, have been averaged with a 1~min time resolution in order to have a uniform time sampling and get rid of the smallest scales fluctuations.
The dataset is split into a faster wind (between 3 and 33\% part of the distribution starting from the largest $u_{p, R}$ value) and slower wind (between 3 and 33\% lowest $u_{p, R}$), see Appendix \ref{sec:appendix_data_selec_wind_speeds} for more details.
After the separation, remaining SolO measurements with $| \mathbf{v} | < 260$~km/s have been discarded due to unreliable temperature measurements of the PAS instrument at lower energies\footnote{see the SWA-PAS L2 data user guide at \url{https://www.cosmos.esa.int/web/soar/instrument-documentation}.} 
Radial averages, with bin resolution of $\Delta R = 10^{-2}$~au for PSP and $\Delta R = 2 \times 10^{-2}$~au for SolO, are shown as plain colored lines, in red for the slower wind and blue for the faster one.
These averages plus and minus the corresponding standard deviation $\sigma( \ln C_{p (\perp, \parallel)})$ are displayed with semi-transparent corresponding colors to indicate the measurements dispersion.

The parallel proton invariant $C_{p \parallel}$ shown in Figure \ref{fig:adiabatic_invariants} (a) exhibits no clear departure from adiabaticity in neither PSP nor SolO measurements.
This is true for both the slower and faster wind populations, especially considering the important standard deviation, mostly due to the $(B/n_p)^2$ term.

On the other hand, Figure \ref{fig:adiabatic_invariants} (b) shows a clear increase of the averaged $\ln C_{p \perp}$ with distance from the Sun, for the two solar wind speed populations.
This increase is significant, even when considering the distribution's standard deviation.
Hence, this indicates the presence of a non-negligible source of perpendicular proton heating during the solar wind expansion.
This result is in agreement with previous studies at larger distances which have identified a perpendicular proton heating, with Helios 1 \& 2 measurements from 0.3 to 1~au by \citet{Marsch_1982_Helios_protons_evolution,Zaslavsky_2023_SW_heating_rate_evaluation} and for PSP encounters 6-9 for $R \in [0.09, 0.75]$~au by \citet{Mozer_2023_PSP_heating_rates}.
The increase is slower for SolO's measurements than it is for PSP's.
We moreover remark the presence of an offset between PSP and SolO data (between 0.25 and 0.3~au), which is most likely due to a difference in the instruments calibration.

\begin{figure}[ht!]
\centering
      \includegraphics[width=\hsize]{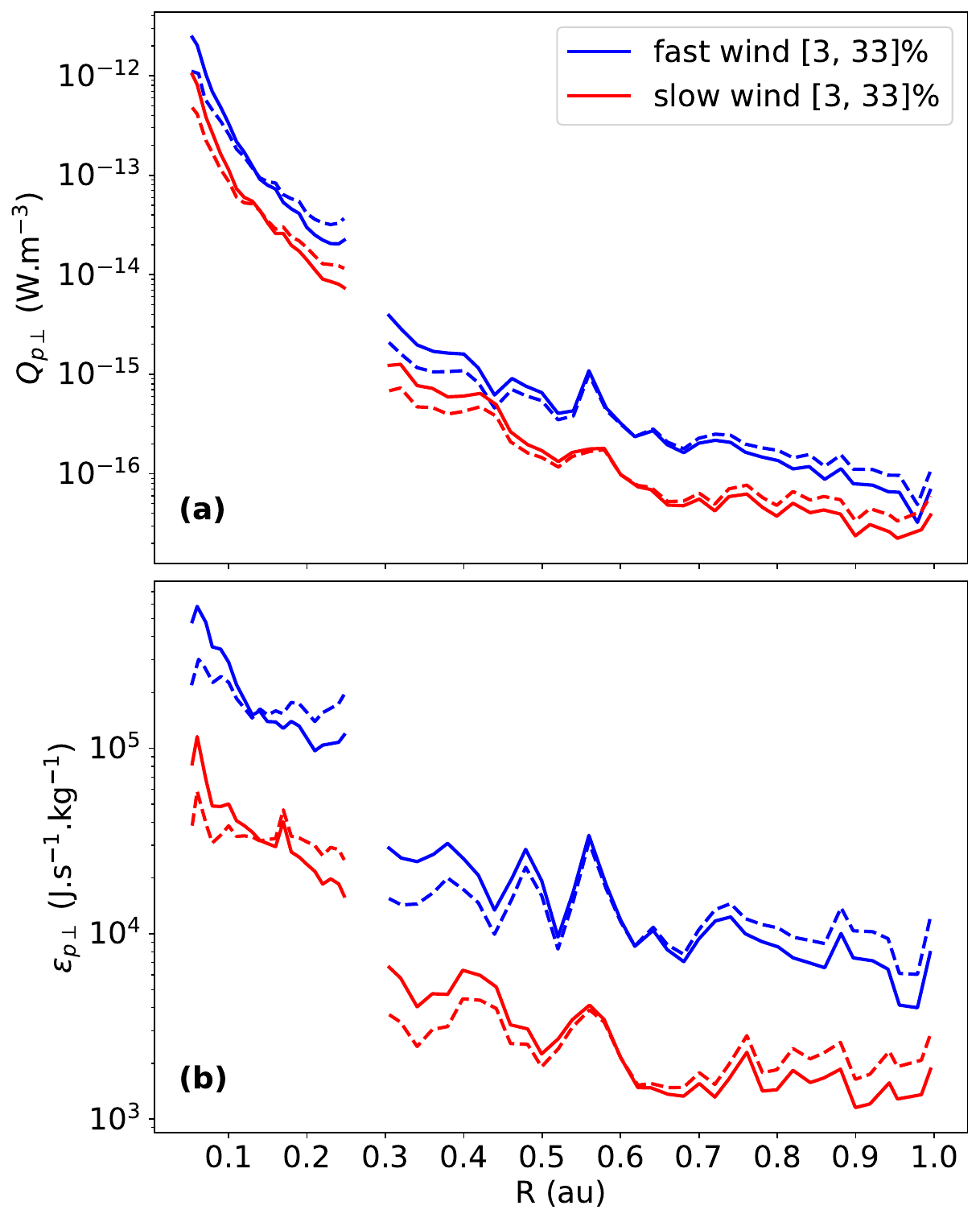}
      \caption{
            Perpendicular non-adiabatic proton heating rate density $Q_{p \perp}$ and heating rate per unit mass $\varepsilon_{p \perp}$.
            We consider the same populations of slower wind (in red) and faster wind (in blue) than for Figure \ref{fig:adiabatic_invariants}.
            The dashed colored curves are the heating rates obtained from a linear fitting of $\ln C_{p \perp}$, and the plain ones from power law fittings.
      }
      \label{fig:perp_heating_rate} 
\end{figure}

Figure \ref{fig:perp_heating_rate} displays the radial evolution of the heating rate density $Q_{p \perp}$ in panel (a), and heating rate per unit mass $\varepsilon_{p \perp} = Q_{p \perp} / (m_p n_p)$ in panel (b), for the previously defined wind populations, in blue for the faster wind and in red for the slower one.
Those were computed using Equation (\ref{eq:C_perp_evol}) and assuming $d/dt = u_{p, R} \, \partial_R$.
The derivative $\partial_R \, \ln C_{p (\perp, \parallel)}$ is evaluated using two fits: linear (dashed lines) and power laws (plain lines).
Fittings are done independently for $R \in [0.05, 0.25]$~au and $R \in [0.3, 1]$~au, from measurements averaged over $10^{-2}$~au for PSP and $2 \times 10^{-2}$~au for SolO (plain lines in Figure \ref{fig:adiabatic_invariants}(b)).
Derivatives are taken from each of the fits results, and the heating rates are finally computed using the parameters $u_{p, R}$, $n_p$ and $T_{p \perp}$, also taken as averages over the same radial bins as previously.
The values of $Q_{p \perp}$ and $\varepsilon_{p \perp}$ estimated with the two different fittings are close and share similar evolution.

The average perpendicular heating rates are higher in the faster wind than the one in the slower one, by a factor around 1.5 to 3 times in $Q_{p \perp}$, and by about a factor 5 in $\varepsilon_{p \perp}$.
The heating rates $Q_{p \perp}$ significantly decrease with distance, mainly due to the plasma spherical expansion inducing a strong decrease of the density with $R$.
There is also an overall decrease with distance in $\varepsilon_{p \perp}$, but to a significantly lesser degree.
Moreover, there are significant shifts (by a factor 5-10 for $\varepsilon_{p \perp}$) between the heating rates (both $Q_{p \perp}$ and $\varepsilon_{p \perp}$) estimated from PSP measurements around 0.25~au, and those estimated from SolO measurements around 0.3~au.
The exact cause of these shifts is not yet fully known, although at least some contribution comes from the $C_{p \perp}$ increases being faster in PSP's than in SolO's data. 
We note that these shifts are always lower for the power law fits estimations (plain curves), which then seem to provide the most relevant heating rate estimation.


\section{Proton Temperature Anisotropy Evolution} \label{sec:anisotropy_evolution}

\subsection{Average temperature anisotropy evolution with distance from the Sun}

\begin{figure}[ht!]
      \centering
      \includegraphics[width=\hsize]{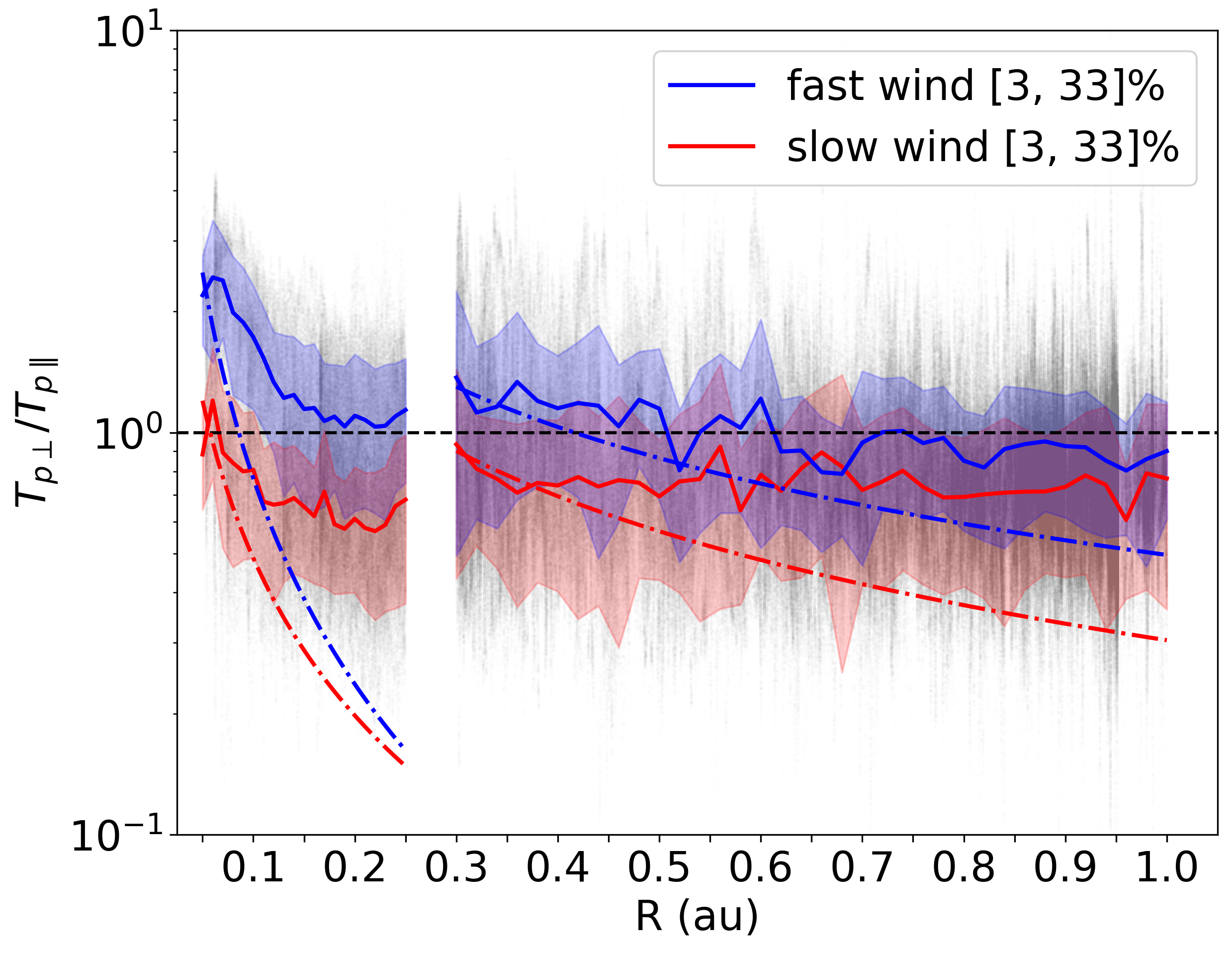}
      \caption{
            Radial evolution of the solar wind proton temperature anisotropy ratio $T_{p \perp} / T_{p \parallel}$ measured by PSP.
            The data and format are the same as in Figure \ref{fig:adiabatic_invariants}.
            The dashed-dotted curves represent the expected evolution from an adiabatic-CGL spherically expanding steady-state plasma.
      }
      \label{fig:T_anisotropy_evol} 
\end{figure}

We are now interested in the effects of the previously identified perpendicular heating on the plasma evolution.
More precisely, we focus on the proton temperature anisotropy which, as mentioned in Section \ref{sec:theory-equations}, would be expected to display a strong decrease of $T_{p \perp} / T_{p \parallel}$ for an adiabatic expansion of the solar wind close to the Sun.

Figure \ref{fig:T_anisotropy_evol} shows the radial evolution of the proton temperature anisotropy $T_{p \perp} / T_{p \parallel}$ measured by PSP and SolO, with the same format as Figure \ref{fig:adiabatic_invariants}.
The two considered wind populations remain well separated throughout the probed radial distances, with the average $T_{p \perp} / T_{p \parallel}$ consistently remaining higher in the faster wind than in the slower one.
A significant decrease is present within PSP's distance range, while the average temperature anisotropies stay weak (close to 1) within SolO's range.
Moreover, the faster wind has averages of $T_{p \perp} / T_{p \parallel} > 1$ for $R < 0.6$~au, while the slower one presents $T_{p \perp} / T_{p \parallel} < 1$, except close to the Sun as $T_{p \perp} / T_{p \parallel} \gtrsim 1$ for $R < 0.075$~au.
As for the adiabatic invariants, there is a shift between the two spacecraft measurements, with SolO near 0.3~au showing on average a larger $T_{p \perp} / T_{p \parallel}$ than PSP around 0.25~au by about a factor 1.3.


\begin{table}[ht!]
\caption{
Power law exponents$^{a}$ considered for the radial evolution of several physical parameters involved in the adiabatic prediction presented in Figures \ref{fig:T_anisotropy_evol}, \ref{fig:evolution_Brazil_PSP} and \ref{fig:Brazil_SolO}.
The parameters evolution are considered separately for the PSP and SolO spacecraft (SC), as well as wind speed populations.
\label{tab:power_law_exponents}
}
\centering
\begin{tabular}{lc cccc}
\hline
SC & Speed & $s(B)$ & $s(n_p)$ & $s(T_{p \perp}/T_{p \parallel})_{\rm CGL}$ & $s(\beta_{p \parallel})_{\rm CGL}$ \\
\hline
  PSP  & Slow & -1.9 & -2.2 & -1.3 & 1.0 \\
  PSP  & Fast & -1.9 & -2.0 & -1.7 & 1.6 \\
  SolO & Slow & -1.7 & -2.1 & -0.9 & 0.5 \\
  SolO & Fast & -1.6 & -2.0 & -0.8 & 0.4 \\
  Radial & -  & -2.0 & -2.0 & -2.0 & 2.0 \\
\hline
\end{tabular}
\tablefoot{$^{a}$ These exponents were obtained through fittings of the magnetic field and densities radial evolution with power-laws $\propto R^s$ and are only here to provide an indicative comparison with a double-adiabatic CGL prediction relevant for the studied dataset.
Therefore, caution must be applied for their interpretation.
Furthermore, these should not be used to derive any other quantities, such as heating rates, as it would result in flawed estimations \citep{Zaslavsky_2023_SW_heating_rate_evaluation}.
}
\end{table}

Table \ref{tab:power_law_exponents} provides, for several parameters $A$, the power law exponents $s(A)$ assuming $A \propto R^{s(A)}$.
The exponents $s(B)$ and $s(n)$ are obtained through fitting of PSP and SolO data with radial distance, while, according to Equation (\ref{eq:anisotropy-beta_evol_adiab}), we consider the adiabatic CGL scaling of proton temperature anisotropy $s(T_{p \perp}/T_{p \parallel})_{\rm CGL} = 3s(B) - 2s(n)$ and beta parallel $s(\beta_{p \parallel})_{\rm CGL} = 3s(n) - 4s(B)$.
For comparison, the last line of Table \ref{tab:power_law_exponents} gives the scalings for a spherically expanding solar wind with constant speed and radial magnetic field, see Section \ref{sec:theory-equations} and Equation (\ref{eq:anisotropy-beta_evol_adiab_R}).
The colored dash-dotted curves in Figure \ref{fig:T_anisotropy_evol} then indicate the steady-state double-adiabatic CGL predictions for the slower (red) and faster (blue) solar wind populations assuming power law decreases of $T_{p \perp} / T_{p \parallel}$ given in Table \ref{tab:power_law_exponents}.
We emphasize that these scalings are only provided in order to compare the proton temperature anisotropy measurements with an approximate double-adiabatic CGL prediction relevant for the studied dataset.
Proper evaluation of the density and magnetic field decreases with distance would require further investigations going beyond the scope of this study.

For PSP which is close the Sun, the obtained magnetic field scaling $B \propto R^{-1.9}$ in Table \ref{tab:power_law_exponents} are, as expected, very close to the purely radial field prediction $B \propto R^{-2}$, while the slow wind density scaling $n_p \propto R^{-2.2}$, which is steeper than $R^{-2}$ and therefore consistent with the average plasma acceleration (see Appendix \ref{sec:appendix_data_selec_wind_speeds}).
Due to the Parker spiral development, the magnetic field magnitude evolution gradually changes with distance from the Sun.
For SolO distance range, this results in a $B \propto R^{-1.6}$ and $B \propto R^{-1.7}$, which agrees with values obtained from Helios measurements \citep{Schwenn-Marsch_1990_Helios_book_I}, and are a bit steeper than the results of \citet{Hanneson_2020_IMF_radial_evol} from in-situ magnetic field measurements from the VEX, MESSENGER, ACE and STEREO spacecraft between $\sim$~0.3 and $\sim$~1~au.
The density decrease are, as for PSP, close to $n_p \propto R^{-2}$, in accordance with a spherical expanding plasma with a weakly evolving velocity.
This finally leads to double adiabatic scalings $s(T_{p \perp} / T_{p \parallel})_{\rm adiab}$ far less abrupt than in PSP's case.

As showed in Figure \ref{fig:T_anisotropy_evol}, there is an overall decrease of the average $T_{p \perp} / T_{p \parallel}$ for both wind populations.
These decreases are however less important than the adiabatic prediction, in particular for PSP, and slow down with distance from the Sun.
This behavior highlights the importance of the identified perpendicular heating.
The decrease of $T_{p \perp} / T_{p \parallel}$ is closer to the adiabatic prediction in SolO's data than in PSP's, as expected from their different increase in $C_{p \perp}$ (Figure \ref{fig:adiabatic_invariants} (b)).

\subsection{Radial evolution of the measurements in the $\beta_{p \parallel} \, \textit{-} \, T_{p \perp} / T_{p \parallel}$ plane}

\begin{figure*}[ht!]
      \centering
      \includegraphics[width=\hsize]{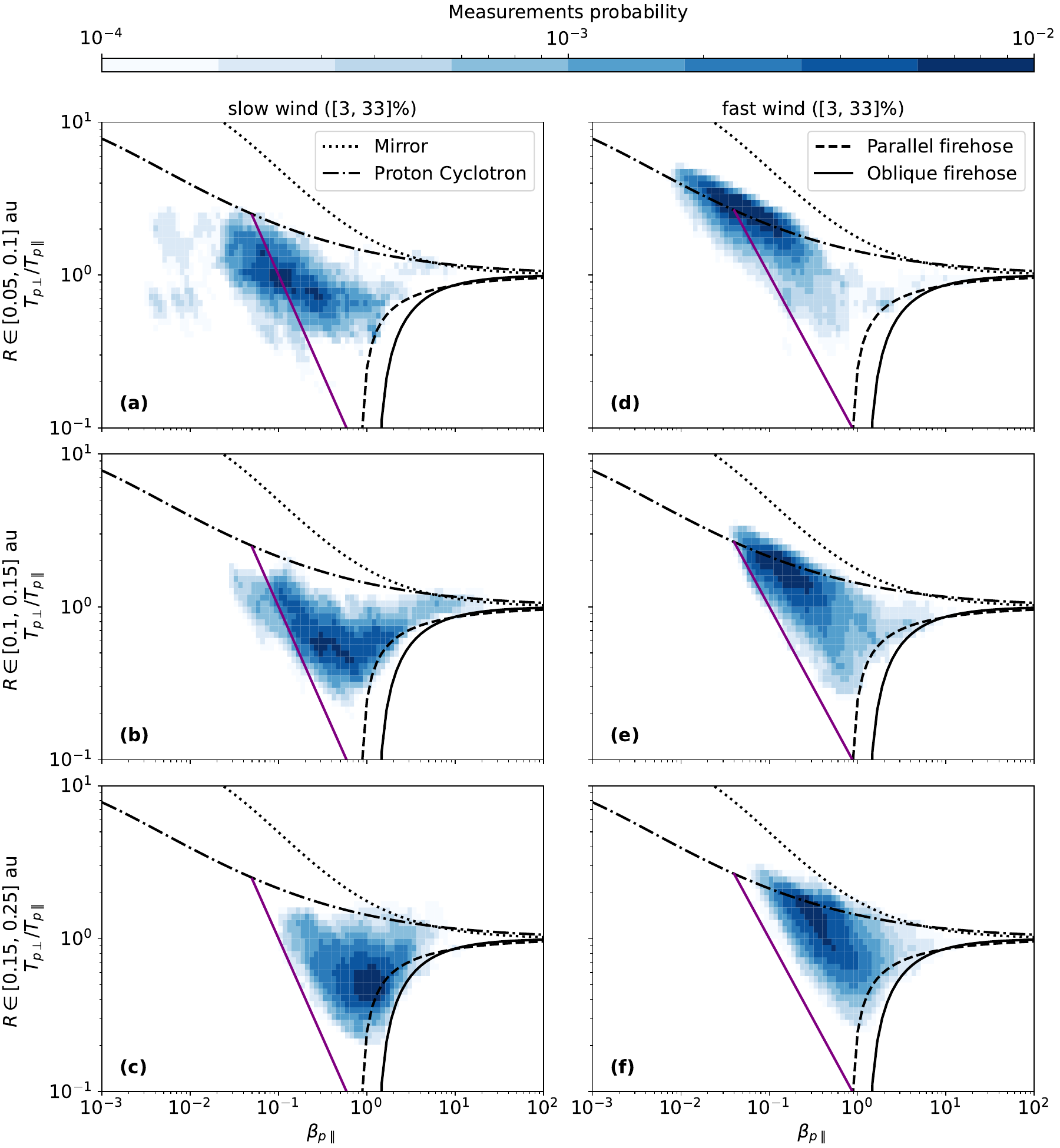}
      \caption{
            Evolution of the solar wind plasma distribution in the $\beta_{p \parallel} \, \textit{-} \, T_{p \perp} / T_{p \parallel}$ plane measured by PSP at distances $R \in [0.05, 0.25]$~au from the Sun.
            Marginal stability conditions, considering a growth rate $\gamma = 10^{-3} \, \Omega_{ci}$, for the parallel and oblique proton firehose instabilities are indicated by the black dashed and plain curves, while those for the proton cyclotron and mirror instabilities are indicated by the dashed-dotted and dotted black curves \citep{Hellinger_2006_anisotropy_thresholds_SW}.
            The considered time intervals are the same as in Figure \ref{fig:adiabatic_invariants} and \ref{fig:T_anisotropy_evol}, but using here measurements at the full time resolution from the SPAN-Ion instrument.
      }
      \label{fig:evolution_Brazil_PSP} 
\end{figure*}

Since temperature anisotropies and their regulation are closely linked with quasi-linear proton instabilities, we now consider the evolution of the plasma distribution in the $\beta_{p \parallel} \, \textit{-} \, T_{p \perp} / T_{p \parallel}$ plane.
We study here the same dataset as for Figures \ref{fig:adiabatic_invariants} and \ref{fig:T_anisotropy_evol}, but without temporal averaging.
Figure \ref{fig:evolution_Brazil_PSP} shows PSP measurements split into three different distances intervals, written in the panels labels. 
As previously we consider two solar wind speed populations, defined here in each of the three radial interval. 
Colors are the measurements' discrete probability, defined here as 2D histograms discretized over $(100 \times 50)$ logarithmically spaced bins in the $\beta_{p \parallel} \, \textit{-} \, T_{p \perp} / T_{p \parallel}$ plane, and normalized by the total number of measurements at the corresponding distance range and wind speed. 
In order to better quantify the temperature anisotropy evolution, the percentage of measurements with $T_{p \perp} / T_{p \parallel} < 1$ corresponding to the data displayed on Figure~\ref{fig:evolution_Brazil_PSP} are moreover given in Table~\ref{tab:percentage_anisotropy}.

Black curves in Figure \ref{fig:evolution_Brazil_PSP} indicate marginal stability conditions of linear kinetic proton instabilities for a growth rate $\gamma = 10^{-3} \, \Omega_{ci}$ \citep{Hellinger_2006_anisotropy_thresholds_SW}.
These conditions are useful to separate stable from largely unstable regions, but caution should be applied in the interpretation as those are derived assuming "ideal" conditions of an homogeneous proton-electron plasma with velocity distribution functions that are bi-Maxwellian for the protons and Maxwellian with $\beta_{e} = 1$ for the electrons.
The double adiabatic CGL predictions according to Table \ref{tab:power_law_exponents} are indicated by the purple line in the different panels of Figure \ref{fig:evolution_Brazil_PSP}.

\begin{table}[ht!]
\caption{
Percentage of measurements with a proton temperature anisotropy $T_{p \perp}/T_{p \parallel} <1$ for both the slow wind and fast wind population, indicated by the subscripts "SSW" and "FSW", respectively.
The data and distances ranges considered are the same as in Figures \ref{fig:evolution_Brazil_PSP} and \ref{fig:Brazil_SolO}.
\label{tab:percentage_anisotropy}
}
\centering
\begin{tabular}{cccc}
\hline
SC & $R$ (au) & $(T_{p \perp}/T_{p \parallel} <1)_{\rm SSW}$ & $(T_{p \perp}/T_{p \parallel} <1)_{\rm FSW}$ \\
\hline
       & [0.05,  0.1] & 64\% & 10\% \\
  PSP  & [0.1,  0.15] & 87\% & 28\% \\
       & [0.15, 0.25] & 92\% & 45\% \\
\hline
       & [0.3,   0.5] & 76\% & 45\% \\
  SolO & [0.5,  0.75] & 79\% & 63\% \\
       & [0.75,    1] & 87\% & 70\% \\
\hline
\end{tabular}
\end{table}

The slower wind distribution undergoes a strong evolution, see Figure \ref{fig:evolution_Brazil_PSP}(a-c).
For $R \in [0.05, 0.1]$~au (panel (a)), the largest part of the distribution is spread out around $\beta_{p \parallel} = 0.1$ and $T_{p \perp} / T_{p \parallel} = 1$ away from the marginal instability conditions, and with an anti-correlation between $\beta_{p \parallel}$ and $T_{p \perp} / T_{p \parallel}$.
At larger distances, for $R \in [0.1, 0.15]$~au (panel (b)), the bulk of the distribution remains spread around a similar diagonal, but has moved toward higher values of $\beta_{p \parallel}$ and smaller values of $T_{p \perp} / T_{p \parallel}$, with a larger part of the measurements now being close to the parallel firehose stability condition (dashed curve). 
Data with the largest $\beta_{p \parallel}$ values are well constrained by the instabilities thresholds within a long narrow corridor.
Then, for $R \in [0.15, 0.25]$~au (panel (c)), more of the distribution has continued migrating towards the larger $\beta_{p \parallel}$ and lower $T_{p \perp} / T_{p \parallel}$ values, while still being constrained by the firehose instabilities thresholds (dashed and plain curves).
An important part of the measurements however remain distributed in the intermediate values of the plane, away from the instability curves.
Furthermore, the distribution evolution deviates noticeably from the adiabatic prediction (in purple), which we attribute to the previously identified perpendicular heating (Figure \ref{fig:perp_heating_rate}).

The distribution and evolution of the faster wind is significantly different than that of the slow wind, see Figure \ref{fig:evolution_Brazil_PSP} (d-f) and Table \ref{tab:percentage_anisotropy}.
Close to the Sun, for $R \in [0.05, 0.1]$~au (panel (d)), the larger part of the distribution is clustered around values of low beta parallel $\beta_{p \parallel} \in [0.02, 0.2]$ and high temperature anisotropy $T_{p \perp} / T_{p \parallel} \in [2, 3]$, near the proton cyclotron instability condition (dashed-dotted curve).
Further from the Sun (panels (e) and (f)), the distribution evolves towards regions of larger $\beta_{p \parallel}$ and lower $T_{p \perp} / T_{p \parallel}$.
Some part of the distribution (smaller than in the slow wind's case) attains and, seems constrained by, the firehose instability conditions.
An important number of measurements remains close to the proton cyclotron curve with a non-negligible part even being in the unstable region.
The overall distribution evolution is, as expected, different from the adiabatic prediction (purple), which is consistent with the identified perpendicular heating.

\begin{figure*}[ht!]
      \centering
      \includegraphics[width=\hsize]{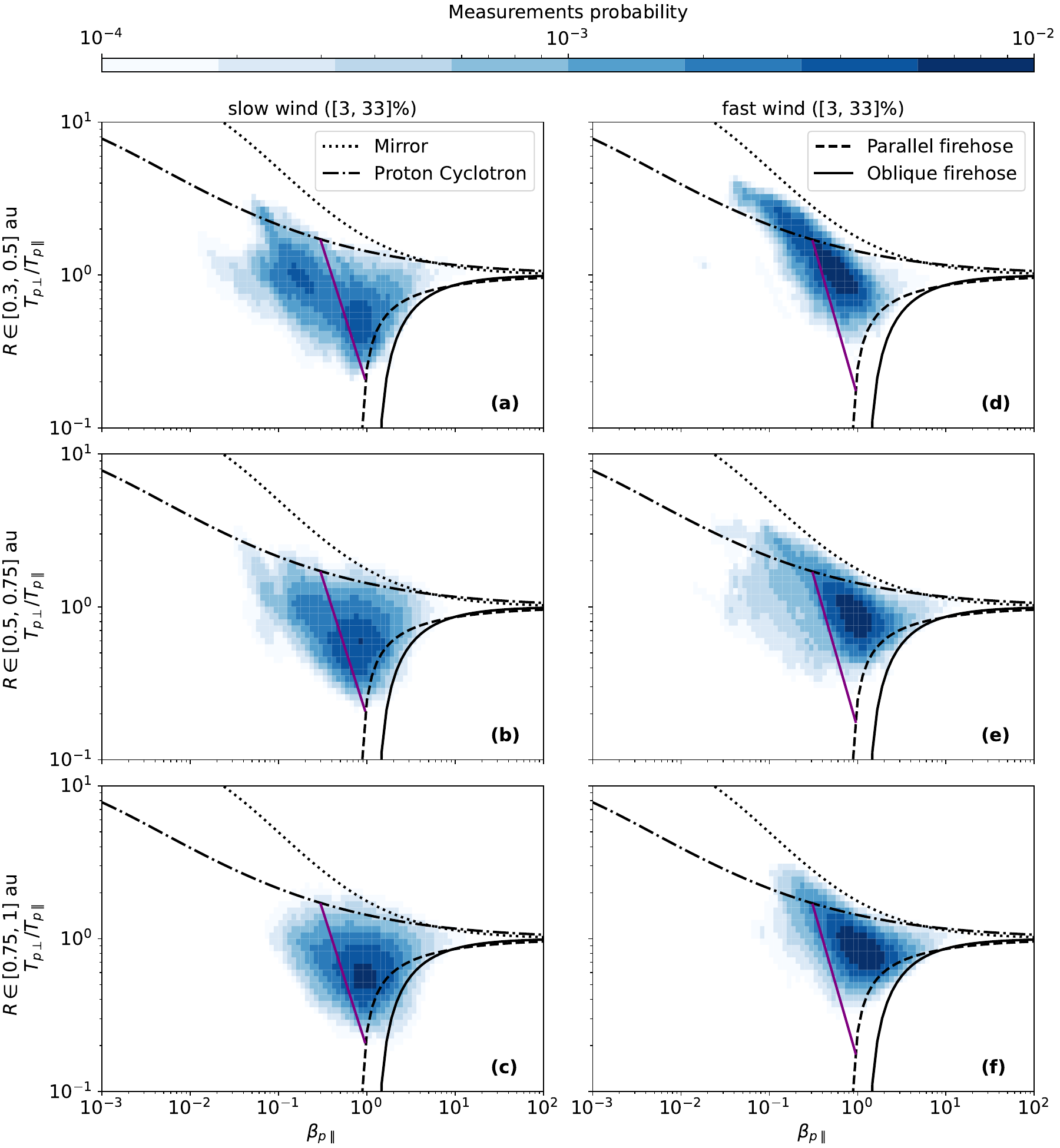}
      \caption{
            Evolution of the solar wind plasma distribution in the $\beta_{p \parallel} \, \textit{-} \, T_{p \perp} / T_{p \parallel}$ plane measured by SolO at distances $R \in [0.3, 1]$~au, with the same format as in Figure~\ref{fig:evolution_Brazil_PSP}.
      }
      \label{fig:Brazil_SolO}
\end{figure*}

Figure \ref{fig:Brazil_SolO} shows the evolution of SolO measurements in the $\beta_{p \parallel} \, \textit{-} \, T_{p \perp} / T_{p \parallel}$ plane.
As in Figure \ref{fig:evolution_Brazil_PSP}, we consider three different distances ranges: panels (a) and (d) give the measurements normalized histograms for distances $R \in [0.3, 0.5]$~au, panels (b) and (e) for $R \in [0.5, 0.75]$~au and panels (c) and (f) for $R \in [0.75, 1]$~au.
Likewise, percentages of measurements with $T_{p \perp} / T_{p \parallel} < 1$ are indicated in Table \ref{tab:percentage_anisotropy} in order to provide a more quantitative idea of the temperature anisotropy evolution.
The purple curves in Figure \ref{fig:Brazil_SolO} indicate the adiabatic prediction based on slow/fast wind fittings with radial distance of the density $n_p$ and magnetic field magnitude $B$ from our dataset, as was done for PSP, see Table \ref{tab:power_law_exponents}.

Measurements are close to, but not completely coinciding with the furthest PSP measurements from the Sun, as could be expected from the observed shift in Figure \ref{fig:T_anisotropy_evol}, see also Table \ref{tab:percentage_anisotropy}.
In the slower wind, the measurements closer to the Sun (panel (a)) are mostly distributed around $\beta_{p \parallel} = 0.5$ and $T_{p \perp} / T_{p \parallel} = 0.7$.
They are bounded by the firehose instabilities and proton cyclotron conditions (black dashed/plain and dashed-dotted curves), and display an anti-correlation between $\beta_{p \parallel}$ and $T_{p \perp} / T_{p \parallel}$.
Then, with distance (panels (b) and (c)), an important part of the distribution remains clustered near the firehose conditions, while the rest seems to spread out in the $\beta_{p \parallel} \, \textit{-} \, T_{p \perp} / T_{p \parallel}$ plane.
We note that PSP slower wind measurements are already distributed close to the firehose instability conditions for $R \in [0.15, 0.25]$~au, see Figure \ref{fig:evolution_Brazil_PSP} (c).

Regarding the faster wind, most of SolO measurements closer to the Sun (Figure \ref{fig:Brazil_SolO} panel (d)) present a stronger anti-correlation between $\beta_{p \parallel}$ and $T_{p \perp} / T_{p \parallel}$, than in the slower wind.
The data are spread from the mirror instability to the firehose instabilities conditions. 
Further away (panels (e) and (f)), the anti-correlation remains while the core of the distribution moves closer to the firehose instability curves, by which it seems constrained.
The overall evolution qualitatively agrees with the Helios observations studied in \citet{Matteini_2007_anisotropy_Helios-Ulysses}.

Finally, the faster wind $T_{p \perp} / T_{p \parallel} > 1$ temperature anisotropies are better constrained by the mirror instability conditions than the proton cyclotron's ones.
However, as mentioned above, the fitted stability conditions rely on several assumptions which are not verified in the solar wind, such as being in a homogeneous plasma with bi-Maxwellian velocity distributions functions.
For example, using fast solar wind Ulysses data between 1.5 and 4~au, \citet{Matteini_2013_core-beam_protons} have shown that the plasma distribution in the $\beta_{p \parallel} \, \textit{-} \, T_{p \perp} / T_{p \parallel}$ plane, as well as its radial evolution, are significantly different if the considered temperatures are resulting from the core, beam, or total proton velocity distribution functions - highlighting the importance of non-Maxwellianity and kinetic features.
Most notably, the proton beam can significantly affect the dispersion properties of the plasma and be driver of instabilities \citep{Daughton-Gary_1998_proton_beam_instab_theory, Gary_2016_instability_SW_beam, Klein_2021_PSP_linear_stability_beam}, principally due to its drift with the proton core, which is on average larger in the faster wind streams than in the slower ones \citep{Marsch_1982_Helios_protons_evolution, Durovcova_2019_Helios_ions_drifts}.
Similarly, an alpha particles population (with or without a velocity drift) can also modify the plasma linear marginal stability conditions \citep{Price_1986_instab_mirror_proton_cyclotron_alpha, Gary_2003_proton-alpha_instab, Hellinger_2005_protons-alphas_instab, Hellinger_2006_proton_FH_alpha, Matteini_2012_ion_kinetics_evol_SW, Martinovic_2026_oblique_drift_instab}, which, when included in the study of \citet{Matteini_2007_anisotropy_Helios-Ulysses} led to Alfvén cyclotron instability thresholds better matching Helios and Ulysses data.
More generally, the contribution from all the different particles population can affect the plasma dispersion properties and therefore modify the linear instability growth rates \citep{Gary_1993_instabilities_book, Dasso_2003_instabs, Chen_2016_multispecies_instab}. 
Additionally, plasma stability conditions are generally derived under a linear (or quasi-linear) framework, which might not be accurately representing the complex solar wind conditions.

\section{Discussion and conclusions} \label{sec:conclusion}

In this study, we investigated the solar wind proton heating and temperature anisotropy evolution with distance from the Sun.
For that, we have analyzed PSP measurements from encounters 1-25 at distances $R \in [0.05, 0.25]$~au, and SolO measurements at $R \in [0.3, 1]$~au.
Since the data used here result from integration of the proton velocity distributions functions, the temperatures are indicative of the total proton internal energy in the different directions.
This includes contributions from the beam, which can be important for $T_{p \parallel}$.
Indeed, for the solar wind with speed $|\mathbf{u}| < 500$~km/s measured by SolO in the distance range [0.3, 1]~au, \citet{Ioannou_2026_SolO_beam_T_anisotropy_evolution} estimated that, on average, the proton beam drift along the local magnetic field accounts for $\sim 40 \%$ of the total parallel proton temperature.
We note that the interpretation of our results regarding PSP measurements might be limited by the SPAN-Ion instrument finite FOV, and the obstruction of incoming particles by PSP's heat shield.
Measurements reliability due to FOV effects start to decrease for $R>0.135$~au ($30\, R_s$) although most measurements principally suffer from it at distances larger than $R = 0.25$~au ($\sim 55 \, R_s$), \citet[see Figure 4 in][]{Yogesh_2026_PSP_radial_evolution}.
We therefore expect our results and interpretations to remain, at least qualitatively, valid despite these instrumental issues.

The measurements have been split into a slow and fast wind populations based on percentiles of the radial solar wind speed \citep[see][and Appendix \ref{sec:appendix_data_selec_wind_speeds}]{Maksimovic_2020_anticorrel_V-Te, Dakeyo_2022_isopoly}. 
Indeed, our dataset indicates an average acceleration of the solar wind with distance from the Sun (and even more so for the slow wind measured within PSP's distance range), with a speed increase of about 150-200~km/s from 0.05 to 1~au for both wind populations, see Appendix \ref{sec:appendix_data_selec_wind_speeds}.
As compared to other studies dividing the solar wind into a faster and slower component based on a constant speed threshold \citep[e.g.][]{Payne_2026_non-adiaba_SBs_PSP}, this method therefore not only allows to account for the average plasma acceleration, but also to have a more reliable separation of the wind populations with distance.

On average, the proton perpendicular invariant $C_{p \perp}$ increases with distance for both the slower and faster solar wind, indicating the presence of non-adiabatic perpendicular heating.
On the other hand, monitoring of the parallel invariant $C_{p \parallel}$ does not reveal any clear large-scale departure from adiabaticity for both wind speed populations.
These results extend, and are consistent with those of \citet{Marsch_1983_adiabatic_invariants_Helios, Perrone_2019_FSW_Helios_adiab_inv, Zaslavsky_2023_SW_heating_rate_evaluation} using Helios measurements ($R \in [0.3, 1]$~au), and the study of \citet{Mozer_2023_PSP_heating_rates} on PSP measurements for encounters 6-9 with $R \in [0.09, 0.75]$~au.

We also provide an estimation of the perpendicular proton heating rates.
These rates decrease with distance, with $Q_{p \perp}$ going from $\sim 10^{-12}$~W.m$^{-3}$ at $R = 0.05$~au to $\sim 5 \times 10^{-17}$~W.m$^{-3}$ at $R = 1$~au, while $\varepsilon_{p \perp}$ evolves from around $10^{5}$~J.s$^{-1}$.kg$^{-1}$ at $R = 0.05$~au to around $5 \times 10^{3}$~J.s$^{-1}$.kg$^{-1}$ at $R = 1$~au.
Both $Q_{p \perp}$ and $\varepsilon_{p \perp}$ are moreover always significantly higher in the faster wind than in the slower wind, by about a factor 1.5-3 for $Q_{p \perp}$, and by around a factor 5 for $\varepsilon_{p \perp}$ (Figure \ref{fig:perp_heating_rate}).
The heating rates densities we derive are in agreement with those obtained by \citet{Mozer_2023_PSP_heating_rates} and the slow wind estimations of \citet{Kontar_2025_KAWs_heating}, but lower (by an order of magnitude) than those of \citet{Bowen_2024_ICWs_heating}. 
Moreover, we derive values of $\varepsilon_{p \perp}$ near the Sun which are compatible with the turbulent energy transfer rate estimated by \citet{Bandyopadhyay_2020_PSP_turbulent_energy_transfer_rate} during PSP's first encounter ($\varepsilon \sim 10^{4} \text{-} 10^{5}$~J~s$^{-1}$~kg$^{-1}$), and with those estimated near 1 au \citep[\text{}$\varepsilon \sim 10^{3} \text{-} 10^{4}$~J~s$^{-1}$~kg$^{-1}$, see][]{Vasquez_2007_energy_transfer_rates_1_au, MacBride_2008_turbulent_cascade_rate_1_au, Coburn_2014_cascade_rate_1au}, see also \citet{Marino_Sorriso-Valvo_2023_energy_transfer_rate_review} and references therein.

The average proton temperature anisotropy ratio $T_{p \perp} / T_{p \parallel}$ decreases with solar distance slower than the adiabatic prediction.
Our results imply that the identified perpendicular heating is important enough to significantly affect the solar wind's temperature anisotropy evolution.
Additionally, in agreement with previous studies between 0.3 and 1~au \citep{Marsch_1982_Helios_protons_evolution, Jagarlamudi_2025_SolO_T_anisotropies} the slower and faster wind distributions have well separated anisotropy averages, with $T_{p \perp} / T_{p \parallel}$ being higher in the faster wind than in the slower one (by a factor decreasing with distance from 2.5 to 1.2).
This is consistent with our estimations of the perpendicular heating rates also being higher in the faster wind.
At distances $R < 0.075$~au, an average $T_{p \perp} / T_{p \parallel} \gtrsim 1$ is observed for both wind populations.
This result agrees with the study of \citet{Yogesh_2026_PSP_radial_evolution}, where the authors found $T_{p \perp} / T_{p \parallel} > 1$ for $R < 0.07$~au in PSP measurements from encounters 1-24, not separated by solar wind speed.
Then, if extrapolated closer to the Sun, this implies that a temperature anisotropy with $T_{p \perp} > T_{p \parallel}$ is also globally expected in the high solar corona.
This provides an observational constraint on the coronal heating mechanism(s) - which are therefore probably preferentially acting in the direction perpendicular to the local magnetic field.

Then, in the parallel proton beta - temperature anisotropy ratio ($\beta_{p \parallel} \, \textit{-} \, T_{p \perp} / T_{p \parallel}$) plane, we find the slow wind distribution to evolve as follows.
For PSP's measurements, the bulk distribution progressively moves towards regions of lower $T_{p \perp} / T_{p \parallel}$ and higher $\beta_{p \parallel}$, with a large portion rapidly being constrained by the firehose instability conditions. 
This evolution is however different from the adiabatic prediction, and although the firehose instability is participating, another phenomenon, such as turbulence, must also be at play as even regions far from the firehose instability thresholds evolve non-adiabatically.
Then, for SolO, the measurements at closest distance to the Sun are consistent with PSP's farthest measurements, but are slightly shifted from them with a more important part near intermediate values of the ($\beta_{p \parallel} \, \textit{-} \, T_{p \perp} / T_{p \parallel}$) plane.
Nonetheless, the distribution continues a similar non-adiabatic evolution, albeit more slowly, while still being constrained by the kinetic firehose and proton cyclotron marginal stability conditions.
There also remains an important spread of the slower wind distribution in the $\beta_{p \parallel} \, \textit{-} \, T_{p \perp} / T_{p \parallel}$ plane at every distance, which suggests that the perpendicular heating is not homogeneous.

The faster wind distribution is overall more clustered around regions of higher $T_{p \perp} / T_{p \parallel}$, and seems bounded by the proton cyclotron / mirror instabilities conditions.
This higher value of the temperature anisotropy ratio is consistent with the perpendicular heating being more important than in the slow wind.
With increasing distance from the Sun, an important part of the distribution evolves near the proton cyclotron and mirror marginal stability conditions.
This suggests that within the faster wind, perpendicular heating may therefore be constrained by these instabilities.
The rest of the faster wind progressively spreads away from the proton cyclotron instability conditions, and part of it appears to be constrained by the proton firehose instabilities conditions.
The faster wind population measured by SolO between $R = 0.3$~au and $R = 1$~au, continues with a similar evolution than the one observed at PSP.
Furthermore, a non-negligible part of the distribution remains beyond the proton-cyclotron marginal stability condition but below the mirror instability's one (by which it seems constrained).
Caution must however be applied in the interpretation because several effects, such as the different particles populations properties, can modify these thresholds, see end of Section \ref{sec:anisotropy_evolution}.
PSP and SolO measurements therefore extend closer to the Sun, and agree with, the study of \citet{Matteini_2007_anisotropy_Helios-Ulysses} using Helios and Ulysses data between 0.3 and 2.5~au, which indicates that with distance from the Sun, the fast solar wind distribution progressively approaches regions of lower $T_{p \perp} / T_{p \parallel}$ and becomes constrained by the firehose instability, while still remaining non-adiabatic.

An extension of this study concerns the solar wind separation by Alfv\'enicity degree, in particular to isolate the alfv\'enic slow wind which presents different characteristics than the rest of the slow wind, and more similarities with the fast wind \citep{Damicis_2021_ASSW_review}.
Preliminary results presented in Appendix \ref{sec:appendix_slow_wind_alfvenicity} seem to indicate that the perpendicular heating is equally efficient in the alfvénic and non-alfvénic slow solar wind. This may imply that similar heating processes are at play in both wind populations. Otherwise, this would indicate that different heating mechanisms add up to give similar heating rates - in which case it would be interesting to seek the physical reasons for this. In any case, further investigations are required in order to draw a more complete picture and better assess the role of alfvénicity regarding the solar wind heating rates.

Then, there remains to identify the mechanisms responsible for the observed heating.
These can include ion cyclotron waves \citep{Bowen_2024_ICWs_heating}, kinetic Alfv\'en waves (KAWs) or KAW structures \citep{Kontar_2025_KAWs_heating}, stochastic heating by low-frequency turbulence \citep{Chandran_2010_stochastic_heating_theory, Chandran_2010_stochastic_heating_CH, Bourouaine_2013_stochastic_heating_Helios, Ervin_2026_diffusive_stochastic_heating}, or more generally the distortion of velocity distribution functions through pressure-strain coupling \citep{Yang_2017_pressure-strain, Hellinger_2022_ion-scale_transition_turbulence_PS} due to small-scale coherent structures generated in turbulence \citep{Mangeney_2001_coherent_structures_solar_wind, Alexandrova_2013_review_turbulence-instabilities, Servidio_2015_kinetic_effects_turbulence, DelSarto_2016_anisotropy_v_shear}.

In conclusion, PSP and SolO measurements revealed the presence of significant perpendicular heating in the solar wind.
This heating is strong enough to affect the temperature anisotropy development, and even more so in the fast wind.
It is also inhomogeneous and decreases with distance from the Sun.
Nonetheless, under the expansion effect, the solar wind measured by PSP still develops temperature anisotropies with $T_{p \perp} / T_{p \parallel} < 1$, which then seemingly get constrained by the firehose instabilities.

\begin{acknowledgements}
The authors thank the anonymous reviewer for the valuable comments and suggestions that helped improve the clarity and quality of this work.
All the authors also thank the Jules Janssen Conference Centre of Paris Observatory for hospitality.
EB acknowledge the Space It Up project funded by the Italian Space Agency, ASI, and the Ministry of University and Research, MUR, under Contract Grant Nos. 2024-5-E.0-CUP and I53D24000060005.
OA acknowledges funding support from CNES.
The authors acknowledge the SWEAP team led by J. Kasper for use of SWEAP data and the FIELDS team for use of FIELDS data (\url{https://doi.org/10.48322/0yy0-ba92}).
The FIELDS experiment on the Parker Solar Probe spacecraft was designed and developed under NASA contract NNN06AA01C.
Solar Orbiter is a mission of international cooperation between ESA and NASA, operated by ESA.
Solar Orbiter SWA data were derived from scientific sensors that were designed and created and are operated under funding provided by numerous contracts from UKSA, STFC, the Italian Space Agency, CNES, the French National Centre for Scientific Research, the Czech contribution to the ESA PRODEX program, and NASA.
Solar Orbiter SWA work at the UCL/Mullard Space Science Laboratory is currently funded by STFC grant ST/X/002152/1.
We are grateful to the teams of the MAG (\url{https://doi.org/10.5270/esa-ux7y320}) and SWA (\url{https://doi.org/10.5270/esa-ahypgn6}) instruments, for providing data used in this study.
\end{acknowledgements}

%

\bibliographystyle{aa} 
\bibliography{references} 

\begin{appendix}
\onecolumn
\section{Internal energy evolution equations} \label{sec:appendix_energy_equations}

The purpose of this Appendix is to remind fundamental equations describing the proton internal energy evolution.
Considered equations and demonstrations can be found with more details in \citet{Hunana_2019_CGL_intro} - to which we only added the collisional contributions (from Vlasov to Boltzmann equation) for completeness.
In the following we omit the subscript "$p$" referring to the protons for simplicity.
The full pressure tensor evolution equation, obtained by integrating the second velocity moment of the Boltzmann equation, writes:
\begin{equation}
      \frac{\partial}{\partial t} \mathbf{P} + \bm{\nabla} \cdot (\bm{u} \mathbf{P} + \mathbf{q}) +  \left\{ \mathbf{P} \cdot \bm{\nabla} \bm{u} + \frac{e}{m} \bm{B} \times \mathbf{P} \right\}^{\rm S} = \mathbf{Q'} \, , \label{eq:pressure_tensor_evol}
\end{equation}
with the symmetric operator of a tensor $\mathbf{A}$ defined as $\left\{ \mathbf{A} \right\}^{\rm S} = \mathbf{A} + \mathbf{A}^{\rm T}$ and $\mathbf{A}^{\rm T}$ the transposed matrix of $\mathbf{A}$.
The cross product between the magnetic field vector $\bm{B}$ and pressure tensor $\mathbf{P}$ is explicitly defined as $(\bm{B} \times \mathbf{P})_{\alpha \beta} = \varepsilon_{i j \alpha} B_i P_{j \beta}$, where $\varepsilon_{i j \alpha}$ is the Levi-Civita symbol.
The proton density $n$, bulk velocity $\bm{u}$, pressure tensor $\mathbf{P}$ and heat flux tensor $\mathbf{q}$ are defined by the moments of the velocity distribution function $f$:
\begin{align}
    n &= \int f \, \mathrm{d}^3 v \, , \\
    \bm{u} &= \frac{1}{n} \int \bm{v} \, f \, \mathrm{d}^3 v \, , \\
    \mathbf{P} &= m \int (\bm{v} - \bm{u}) (\bm{v} - \bm{u}) \, f \, \mathrm{d}^3 v \, , \\
    \mathbf{q} &= m \int (\bm{v} - \bm{u}) (\bm{v} - \bm{u}) (\bm{v} - \bm{u}) \, f \, \mathrm{d}^3 v \, .
\end{align}
The term $\mathbf{Q'}$ corresponds to the collisional heat rate tensor
\begin{equation}
      \mathbf{Q'} = m \int \left( \frac{\partial f}{\partial t} \right)_c (\bm{v} - \bm{u}) (\bm{v} - \bm{u}) \, \mathrm{d}^3 v \, ,
\end{equation}
with $\left( \partial f / \partial t \right)_c$, the temporal evolution of the proton velocity distribution function due to collisions.

Projecting Equation (\ref{eq:pressure_tensor_evol}) in the directions parallel and orthogonal to the magnetic field (with unit vector $\bm{b} = \bm{B} / B$), gives two equations describing the temporal evolution of the parallel pressure $P_\parallel = \mathbf{P} : \bm{b} \bm{b}$ and perpendicular pressure $P_\perp = (\text{Tr} (\mathbf{P}) - \mathbf{P} : \bm{b} \bm{b}) /2$
\begin{align}
    & \frac{\partial}{\partial t} P_{\parallel} + \bm{\nabla} \cdot (P_{\parallel} \bm{u})
    + 2 \left( P_{\parallel} \bm{\nabla} \bm{u} + \mathbf{P}_{ng} \cdot \bm{\nabla} \bm{u} + \frac{1}{2} \bm{\nabla} \cdot \mathbf{q} \right) : \bm{b} \bm{b}
    - \bm{b} \cdot \mathbf{P}_{ng} \cdot \frac{2}{B} \frac{\mathrm{d} \bm{B}}{\mathrm{d} t}
    = \mathbf{Q'}_\parallel \, , \label{eq:pressure_para_evol_appendix} \\
    & \frac{\partial}{\partial t} P_{\perp} + \bm{\nabla} \cdot (P_{\perp} \bm{u}) + P_{\perp} \bm{\nabla} \cdot \bm{u}
    - \left( P_{\perp} \bm{\nabla} \bm{u} + \mathbf{P}_{ng} \cdot \bm{\nabla} \bm{u} + \frac{1}{2} \bm{\nabla} \cdot \mathbf{q} \right) : \bm{b} \bm{b}
    + \bm{b} \cdot \mathbf{P}_{ng} \cdot \frac{1}{B} \frac{\mathrm{d} \bm{B}}{\mathrm{d} t}
    + \text{Tr} \left( \mathbf{P}_{ng} \cdot \bm{\nabla} \bm{u} + \frac{1}{2} \bm{\nabla} \cdot \mathbf{q} \right)
    = \mathbf{Q'}_\perp \, , \label{eq:pressure_perp_evol_appendix}
\end{align}
where the proton pressure tensor $\mathbf{P}$ has moreover been decomposed into two parts: a gyrotropic one $\mathbf{P}_g = P_{\perp} \mathbf{I} + (P_{\parallel} - P_{\perp}) \bm{b} \bm{b} $ and a non-gyrotropic one $\mathbf{P}_{ng} = \mathbf{P} - \mathbf{P}_{g}$ ; we also defined $\mathbf{Q'}_\parallel = \mathbf{Q'} : \bm{b} \bm{b}$ and $\mathbf{Q'}_\perp = \left(\text{Tr} (\mathbf{Q'}) -  \mathbf{Q'}_\parallel \right) /2$.

After some manipulations, Equations (\ref{eq:pressure_para_evol_appendix}) and (\ref{eq:pressure_perp_evol_appendix}) can be written as \citep[see Equations (2.100) and (2.101) in][]{Hunana_2019_CGL_intro}:
\begin{align}
    \frac{1}{2} \frac{\mathrm{d}}{\mathrm{d}t} C_{\parallel} =& \, \frac{C_{\parallel}}{P_{\parallel}} \left[ P_{\parallel} \bm{\nabla} \cdot \bm{u}
    - \left( P_{\parallel} \bm{\nabla} \bm{u} + \mathbf{P}_{ng} \cdot \bm{\nabla} \bm{u} + \frac{1}{2} \bm{\nabla} \cdot \mathbf{q} \right) : \bm{b} \bm{b}
    + \left( P_{\parallel} \bm{b} + \bm{b} \cdot \mathbf{P}_{ng} \right) \cdot \frac{1}{B} \frac{\mathrm{d} \bm{B}}{\mathrm{d} t} + \mathbf{Q'}_\parallel /2 \right], \label{eq:nrj_para_evol_appendix} \\
    \frac{\mathrm{d}}{\mathrm{d}t} C_{\perp} =& \, \frac{C_{\perp}}{P_{\perp}} \left[ - P_{\perp} \bm{\nabla} \cdot \bm{u}
    + \left( P_{\perp} \bm{\nabla} \bm{u} + \mathbf{P}_{ng} \cdot \bm{\nabla} \bm{u} + \frac{1}{2} \bm{\nabla} \cdot \mathbf{q} \right) : \bm{b} \bm{b}
    - \left( P_{\perp} \bm{b} + \bm{b} \cdot \mathbf{P}_{ng} \right) \cdot \frac{1}{B} \frac{\mathrm{d} \bm{B}}{\mathrm{d} t} \right. \nonumber \\
    & \, - \left. \text{Tr} \left( \mathbf{P}_{ng} \cdot \bm{\nabla} \bm{u} + \frac{1}{2} \bm{\nabla} \cdot \mathbf{q} \right) + \mathbf{Q'}_\perp \right] \, , \label{eq:nrj_perp_evol_appendix}
\end{align}
with $C_{\parallel} = T_{\parallel} \left( B / n \right)^2$, and $C_{\perp} = T_{\perp} / B$ the parallel and perpendicular adiabatic invariants, respectively.
These equations can then simply be rewritten as
\begin{align}
    \frac{1}{2}  \frac{\mathrm{d}}{\mathrm{d}t} \ln C_{\parallel} &= Q_{\parallel} / P_{\parallel} \, , \label{eq:C_para_evol_appendix} \\
     \frac{\mathrm{d}}{\mathrm{d}t} \ln C_{\perp} &= Q_{\perp} / P_{\perp} \, . \label{eq:C_perp_evol_appendix}
\end{align}
where $Q_{\perp}$ and $Q_{\parallel}$ are non-adiabatic heating rates and the total non-adiabatic heating rate is then $Q = Q_\parallel + Q_\perp$.
The rates $Q_{p \parallel}$ and $Q_{p \perp}$ include the effects of collisions, heat fluxes, non-ideal electric fields ($\mathbf{E} \neq -\mathbf{u} \times \mathbf{B}$), and contributions from the non-gyrotropic part of the pressure tensor.

We now demonstrate how considering some simplifying assumptions progressively leads to conservation of $C_{\parallel}$ and $C_{\perp}$.
Neglecting heat fluxes $\mathbf{q}$ and collisions effects $\mathbf{Q'}$, Equations (\ref{eq:nrj_para_evol_appendix}) and (\ref{eq:nrj_perp_evol_appendix}) simplify to:
\begin{align}
    \frac{1}{2} \frac{\mathrm{d}}{\mathrm{d}t} C_{\parallel} &= \frac{C_{\parallel}}{P_{\parallel}} \left[ P_{\parallel} \bm{\nabla} \cdot \bm{u}
    - \left( P_{\parallel} \bm{\nabla} \bm{u} + \mathbf{P}_{ng} \cdot \bm{\nabla} \bm{u} \right) : \bm{b} \bm{b}
    + \left( P_{\parallel} \bm{b} + \bm{b} \cdot \mathbf{P}_{ng} \right) \cdot \frac{1}{B} \frac{d\bm{B}}{dt} \right] \, , \label{eq:nrj_para_evol_simple1_appendix} \\
    \frac{\mathrm{d}}{\mathrm{d}t} C_{\perp} &= \frac{C_{\perp}}{P_{\perp}} \left[ - P_{\perp} \bm{\nabla} \cdot \bm{u}
    + \left( P_{\perp} \bm{\nabla} \bm{u} + \mathbf{P}_{ng} \cdot \bm{\nabla} \bm{u}  \right) : \bm{b} \bm{b} 
    - \left( P_{\perp} \bm{b} + \bm{b} \cdot \mathbf{P}_{ng} \right) \cdot \frac{1}{B} \frac{d\bm{B}}{dt}
    - \text{Tr} \left( \mathbf{P}_{ng} \cdot \bm{\nabla} \bm{u}  \right) \right] \, . \label{eq:nrj_perp_evol_simple1_appendix}
\end{align}

Then, assuming ideal Ohm's law ($\bm{E} = - \bm{u} \times \bm{B}$, where inertia of species other than protons have been neglected), the induction equation writes
\begin{equation}
      \frac{\mathrm{d} \bm{B}}{\mathrm{d} t} = - \bm{B} (\bm{\nabla} \cdot \bm{u}) + (\bm{B} \cdot \bm{\nabla}) \bm{u} \, ,
\end{equation}
projecting the previous equation along $\bm{b}$, and reminding that $\bm{b} \cdot \bm{\nabla} \bm{u} \cdot \bm{b} = \bm{\nabla} \bm{u} : \bm{b} \bm{b}$, yields
\begin{equation}
      \bm{\nabla} \bm{u} : \bm{b} \bm{b} = \bm{b} \cdot \frac{1}{B} \frac{\mathrm{d} \bm{B}}{\mathrm{d} t} + \bm{\nabla} \cdot \bm{u} \, .
\end{equation}
This can be used to further simplify Equations (\ref{eq:nrj_para_evol_simple1_appendix}) and (\ref{eq:nrj_perp_evol_simple1_appendix}) into:
\begin{align}
    \frac{1}{2} \frac{\mathrm{d}}{\mathrm{d}t} C_{\parallel} &= \frac{C_{\parallel}}{P_{\parallel}} \left[
    - \left( \mathbf{P}_{ng} \cdot \bm{\nabla} \bm{u} \right) : \bm{b} \bm{b} + \left(\bm{b} \cdot \mathbf{P}_{ng} \right) \cdot \frac{1}{B} \frac{\mathrm{d} \bm{B}}{\mathrm{d} t} \right] \, , \label{eq:nrj_para_evol_simple2_appendix} \\
    \frac{\mathrm{d}}{\mathrm{d}t} C_{\perp} &= \frac{C_{\perp}}{P_{\perp}} \left[
    \left( \mathbf{P}_{ng} \cdot \bm{\nabla} \bm{u}  \right) : \bm{b} \bm{b} - \left( \bm{b} \cdot \mathbf{P}_{ng} \right) \cdot \frac{1}{B} \frac{\mathrm{d} \bm{B}}{\mathrm{d} t}
    - \text{Tr} \left( \mathbf{P}_{ng} \cdot \bm{\nabla} \bm{u}  \right) \right] \, , \label{eq:nrj_perp_evol_simple2_appendix} \\
\end{align}
Finally, assuming the proton pressure tensor to be gyrotropic ($\mathbf{P}_{ng} = 0$) leads to the classic double-CGL \citep{Chew_1956_CGL_invariants} equations:
\begin{align}
    \frac{\mathrm{d}}{\mathrm{d}t} C_{\parallel} &= 0 \, , \label{eq:C_para_adiab_appendix} \\
    \frac{\mathrm{d}}{\mathrm{d}t} C_{\perp} &= 0 \, . \label{eq:C_perp_adiab_appendix}
\end{align}

\section{Fast and slow wind separation} \label{sec:appendix_data_selec_wind_speeds}

\begin{figure}[ht!]
      \sidecaption
      \includegraphics[width=0.7\hsize]{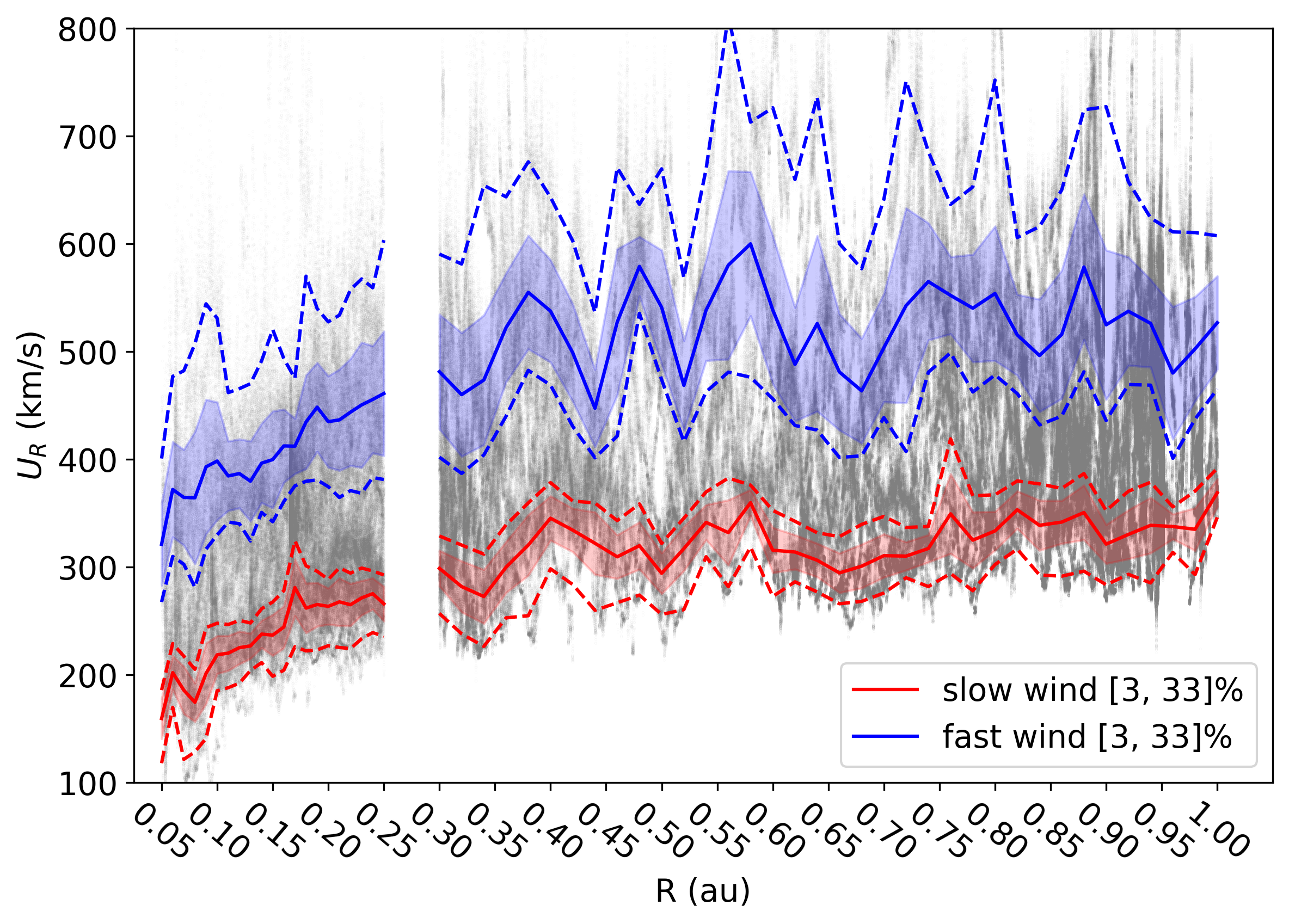}
      \caption{
            Radial evolution of $u_{p, R}$, the proton radial speed.
            The observation times and format are the same as in Figures \ref{fig:adiabatic_invariants} and \ref{fig:T_anisotropy_evol}.
            The colored dashed curves indicate the upper and lower thresholds (3 and 33\%) used for defining the solar wind populations.
      }
      \label{fig:SW_speed_selection}
\end{figure}

In order to account for the general solar wind acceleration with distance, we choose to split the solar wind into different populations based on quantiles of the wind speed distribution, similarly to what done in \citet{Maksimovic_2020_anticorrel_V-Te} and \citet{Dakeyo_2022_isopoly}.
More precisely, for each radial bin of $10^{-2}$~au in PSP measurements and $2 \times 10^{-2}$~au in SolO's, we define a faster wind as the one having a radial speed between the 3\% and 33\% largest $u_{p, R}$, and the slower one with the 3 and 33\% lowest $u_{p, R}$.
The upper and lower bounds of 3\% have been chosen in order to eliminate outliers and thus get a more robust estimation of the averages and standard deviations of computed quantities.
The 3-33\% thresholds taken here allow for both a good separation of the slower and faster wind, while still retaining an important fraction of the total measurement distribution. 

Figure \ref{fig:SW_speed_selection} displays the solar wind speed evolution for the two populations, blue for the faster wind, and red for the slower one.
As Figures \ref{fig:adiabatic_invariants} and \ref{fig:T_anisotropy_evol}, the plain lines give the average values, and the semi-transparent color are the associated standard deviations.
The colored dashed lines show the upper and lower bounds associated with the defined populations.
There is a clear average acceleration of the slower wind population, going from $u_{p, R} \sim 150$-$200$~km/s to $u_{p, R} \sim 350$~km/s.
The faster wind population also displays an average acceleration (from $\sim 350$~km/s to $\sim 500$~km/s), although mostly from 0.125 to 0.25~au and with larger local variations than for the slow wind, especially for $R > 0.3$~au.

\section{Influence of alfvénicity on the slow wind CGL invariants evolution} \label{sec:appendix_slow_wind_alfvenicity}

\begin{figure*}[ht!]
      \centering
      \includegraphics[width=\hsize]{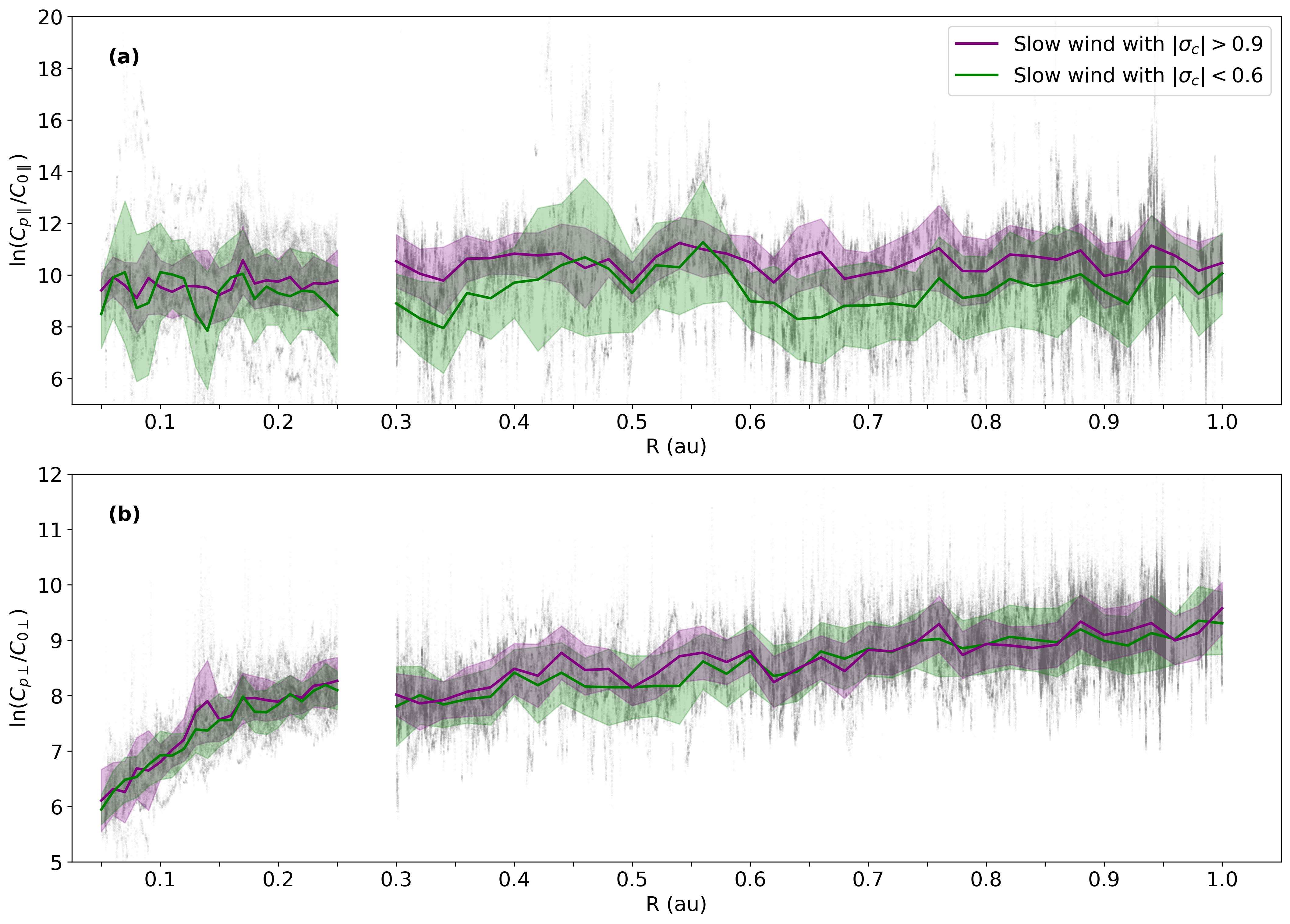}
      \caption{
            Radial evolution of the proton adiabatic invariants depending on the alfvénicity degree in the slow solar wind (see Section \ref{sec:appendix_data_selec_wind_speeds} for the wind speed selection).
            Local averages and standard deviations of the alfvénic slow wind  measurements (with $\sigma_c| > 0.9$) are in purple, and those of the non-alfvénic slow wind measurements (with $\sigma_c| < 0.6$) are in green.
            Otherwise the format is the same as in Figure \ref{fig:adiabatic_invariants}.
      }
      \label{fig:invariants_alfvenicity_slow_wind}
\end{figure*}

A recent growing body of work has focused on a type of slow solar wind presenting a high degree of alfvénicity: the alfvénic slow wind \citep[see e.g.][]{Damicis_2021_ASSW_review, Perrone_2023_ion_kinetics_ASSW, Damicis_2025_SolO_ASSW, Huang_2025_ASSW_anisotropy_evolution}.
A common way to evaluate the alfvénicity content of a solar wind stream is through computation of the reduced magnetic cross helicity $\sigma_c$, defined at each time $t$, and for fluctuations on a scale $\tau$, as
\begin{equation}
    \sigma_c (t, \tau) = \frac{(\delta \bm{z}^+ (t, \tau))^2  - (\delta \bm{z}^- (t, \tau))^2}{(\delta \bm{z}^+ (t, \tau))^2 + (\delta \bm{z}^- (t, \tau))^2 } ,
\end{equation}
where the Els\"asser variables are defined as $\delta \bm{z}^\pm (t, \tau) = \delta \bm{u} (t, \tau) \pm \delta \bm{v}_A (t, \tau)$, with $\bm{v}_A = \bm{B} / \sqrt{m_p n_p \mu_0}$.
There are several possibilities for decomposing fluctuations over scales.
Here, for simplicity, we chose to define fluctuations of each quantity $a(t)$ by subtracting its running average: $\delta a (t, \tau) = a(t) - \langle a(t) \rangle_\tau$ ; we note that with this definition $\delta a (t, \tau)$ includes all fluctuations with time scales smaller than $\tau$.
We set $\tau = 30$~min for consistency with previous studies \citep[e.g.][]{Damicis_2021_first_SolO_ASSW, Perrone_2023_ion_kinetics_ASSW}, although it does not seem this choice of $\tau$  qualitatively influences the results presented below as long as it remains around a few minutes or hours.
In order to investigate if the slow alfvénic wind presents differences of solar wind heating during its evolution, we then separate the slower wind population (see Section \ref{sec:appendix_data_selec_wind_speeds}) into two other groups:  one with with $|\sigma_c| > 0.9$ (alfvénic slow wind) and the other with $| \sigma_c < 0.6|$ (non-alfvénic slow wind).

The proton adiabatic invariants radial evolution of these two alfvénicity groups are displayed in Figure \ref{fig:invariants_alfvenicity_slow_wind}, in purple for $|\sigma_c| > 0.9$, and in green for $|\sigma_c| < 0.6$.
Figure \ref{fig:invariants_alfvenicity_slow_wind} (a) shows that there is no evidence of an average non-adiabatic proton heating or cooling in the direction parallel to $\mathbf{B}$ for neither the alfvénic and non-alfvénic slow wind populations.
Then, on Figure \ref{fig:invariants_alfvenicity_slow_wind} (b) we can see that both the alfvénic and non-alfvénic slow wind have very close values of $C_{p \perp}$ regardless of $R$.
Values of $P_{p \perp}$ are moreover similar for both slow solar wind types (with their ratio $\sim 1$, not shown), which implies similar heating rates in the direction perpendicular to the local magnetic field, see Equation (\ref{eq:C_perp_evol}).
This could be an indication of similar heating mechanisms in the alfvénic and non-alfvénic slow solar winds.

There is, however, a notable difference between the alfvénic and alfvénic slow wind: the invariants' local variations are overall more important in the less alfvénic slow wind, in particular for $C_{p, \parallel}$.
Indeed, the invariants logarithm's standard deviations are larger for the less alfvénic wind, by a factor $\sim 1.3$ for $\ln C_{p, \perp}$ and a factor $\sim 2$ for $\ln C_{p, \parallel}$.
Moreover, the local average of $\ln C_{p, \parallel}$ (Figure \ref{fig:invariants_alfvenicity_slow_wind} (a)) exhibits important local variations with $R$, which are also visible on Figure \ref{fig:adiabatic_invariants} (a).
These variations can probably be explained by crossings of the heliospheric current sheet (HCS) and heliospheric plasma sheet (HPS), which typically exhibit different properties than the rest of the solar wind as well as a lesser alfvénicity degree in their vicinity \citep{Borrini_1981_HCS_superposed_epoch, Winterhalter_1994_HPS, Foullon_2009_layered_HCS, Lavraud_2020_HCS-HPS_PSP, Liou_2021_HCS_superposed_epoch, Shi_2022_HCS_alfvenicity_evolution, Berriot_2025, Fargette_2026_HCS_properties_PSP}.
More specifically, the HCS and HPS typically present a local increase of density and a depletion of magnetic field magnitude, which therefore importantly impact the parallel invariant local variations since $C_{p, \parallel} \propto (B/n_p)^2$.
The magnetic sector boundary proximity (HCS/HPS) could also explain why the average $C_{p, \parallel}$ is generally smaller in the less alfvénic slow wind.

We emphasize that the results presented in this appendix are preliminary and require other investigations, going beyond the scope of the present study, in order to provide a more complete point of view of the influence of alfvénicity on the solar wind heating.
In particular the radial evolution of the slow wind's alfvénicity degree, which is it typically higher close to the Sun, should be characterized.
There is also an important link with the evolution of various kinetic features in the plasma, as a more accurate estimation of the Alfvén velocity contains contribution from the different species's pressure anisotropy and velocity drifts between them \citep{Parker_1958_firehose, Barnes_1971_alfven_waves_generalized_disp_relation, Belcher_1971_large_amplitudes_alfven_waves, Alterman_2018_alpha_proton_beam_drifts}. 
We note, however, that considering other values of the fixed cross-helicity thresholds. (e.g. $|\sigma_c| > 0.8$ or $|\sigma_c| < 0.4$), or using thresholds based on percentiles of the solar wind measurements as for the slow/fast separation, does not seem to lead to results qualitatively departing from those shown in Figure \ref{fig:invariants_alfvenicity_slow_wind}.

\end{appendix}
\end{document}